\documentclass[prd,english,preprintnumbers,amsmath,amssymb,nofootinbib,superscriptaddress,twocolumn]{revtex4}

\pdfoutput=1

\usepackage{mathtools}
\usepackage{amsfonts}
\usepackage{mathrsfs}
\usepackage{bbm}
\usepackage{slashed}

\usepackage{graphicx}
\usepackage[dvipsnames]{xcolor}
\usepackage{array}

\usepackage{xspace}
\usepackage{siunitx}
\usepackage{hyperref}

\usepackage{xifthen}
\usepackage{dsfont}
\usepackage{appendix}
\usepackage{enumitem}
\usepackage{booktabs}
\usepackage{units}

\usepackage[latin1]{inputenc}
\usepackage[dvipsnames]{xcolor}
\usepackage{array}

\setkeys{Gin}{width=0.48\textwidth}

\graphicspath{
	{./}
}

\def\0#1#2{\frac{#1}{#2}}

\def\s0#1#2{\mbox{\small{$ \frac{#1}{#2} $}}}

\newcommand{\I}{\mathrm{i}}
\newcommand{\be}{\begin{eqnarray}}
\newcommand{\ee}{\end{eqnarray}}

\newcommand{\nn}{\nonumber }

\newcommand{\beq}{\begin{equation}}
\newcommand{\eeq}{\end{equation}}
\newcommand{\bea}{\begin{eqnarray}}
\newcommand{\eea}{\end{eqnarray}}
\newcommand{\Nc}{N_{\rm c}}

\def\0#1#2{\frac{#1}{#2}}

\hypersetup{
	colorlinks,
	linkcolor={red!75!black},
	citecolor={blue!75!black},
	urlcolor={blue!75!black},
	pdftitle={Symmetric nuclear matter constraints from the strong interaction},
	pdfauthor={Leonhardt, Pospiech, Schallmo, Braun, Drischler, Hebeler, Schwenk},
	bookmarksopen=true,
	bookmarksopenlevel=2,
	bookmarksnumbered=true
}

\usepackage{babel}
\makeatother
\begin{document}

\title{Symmetric nuclear matter from the strong interaction}

\author{M. Leonhardt} 
\affiliation{Institut f\"ur Kernphysik, Technische Universit\"at Darmstadt, 
D-64289 Darmstadt, Germany}
\author{M. Pospiech} 
\affiliation{Institut f\"ur Kernphysik, Technische Universit\"at Darmstadt, 
D-64289 Darmstadt, Germany}
\author{B. Schallmo} 
\affiliation{Institut f\"ur Kernphysik, Technische Universit\"at Darmstadt, 
D-64289 Darmstadt, Germany}
\author{J. Braun}
\affiliation{Institut f\"ur Kernphysik, Technische Universit\"at Darmstadt, 
D-64289 Darmstadt, Germany}
\affiliation{ExtreMe Matter Institute EMMI, GSI, Planckstra{\ss}e 1, D-64291 Darmstadt, Germany}
\author{C. Drischler}
\affiliation{Department of Physics, University of California, Berkeley, CA 94720}
\affiliation{Lawrence Berkeley National Laboratory, Berkeley, CA 94720}
\author{K. Hebeler}
\affiliation{Institut f\"ur Kernphysik, Technische Universit\"at Darmstadt, 
D-64289 Darmstadt, Germany}
\affiliation{ExtreMe Matter Institute EMMI, GSI, Planckstra{\ss}e 1, D-64291 Darmstadt, Germany}
\author{A. Schwenk}
\affiliation{Institut f\"ur Kernphysik, Technische Universit\"at Darmstadt, 
D-64289 Darmstadt, Germany}
\affiliation{ExtreMe Matter Institute EMMI, GSI, Planckstra{\ss}e 1, D-64291 Darmstadt, Germany}
\affiliation{Max-Planck-Institut f\"ur Kernphysik, Saupfercheckweg 1, 69117 Heidelberg, Germany}

\begin{abstract}
  We study the equation of state of symmetric nuclear matter at zero
  temperature over a wide range of densities using two complementary
  theoretical approaches. At low densities up to twice nuclear
  saturation density, we compute the energy per particle based on modern
  nucleon-nucleon and three-nucleon interactions derived within
  chiral effective field theory. For higher densities we derive 
  for the first time constraints in a Fierz-complete
  setting directly based on quantum chromodynamics
  using functional renormalization group techniques. We find remarkable
  consistency of the results obtained from both approaches as they come together in density 
  and the natural emergence of a
  maximum in the speed of sound $c_S$ at supranuclear densities. 
  The presence of this maximum
  appears tightly connected to the formation of a diquark gap.
  Notably, this maximum is observed to exceed the asymptotic value $c_S^2 = 1/3$ 
  while its exact position in terms of the density cannot yet be determined conclusively. 
\end{abstract}
\maketitle
%
{\it Introduction.--}
The theoretical understanding of the equation of state (EOS) of dense matter
has been one of the main frontiers in nuclear physics in recent decades. While
the EOS of cold matter up to around nuclear saturation density, $n_0 = 0.16
\: \text{fm}^{-3}$, can be constrained by properties of atomic
nuclei~\cite{Tsang:2012se,Lattimer:2012xj,Hebeler:2015hla,Roca-Maza:2018ujj}, the composition and properties of
matter at supranuclear densities as it exists, e.g., in the center of neutron
stars are still open questions. Recent breakthroughs like the first detection of the gravitational wave 
signal of the neutron star merger~\cite{TheLIGOScientific:2017qsa,Abbott:2018wiz} as well as ongoing
missions aiming at first direct neutron star radius measurements using x-ray
timing~\cite{Watts:2016uzu,NICER,NICER2} can significantly enhance our
theoretical understanding of neutron-rich matter under extreme
conditions. Combining information from
these ongoing efforts with existing observational data like precise mass
measurements of heavy neutron stars~\cite{Demorest10,Antoniadis13,Fonseca2016,2019arXiv190406759C}
or also heavy ion collisions~\cite{Danielewicz:2002pu} can
provide further constraints for the EOS. However, all such
measurements can only provide indirect insight into the microscopic nature of
matter at high densities~\cite{Steiner:2010fz,Hebeler:2013nza,Ozel:2016oaf}.
The present work aims to constrain properties of symmetric nuclear matter 
from calculations based on strong interactions with controlled uncertainties, 
without taking into account electromagnetic or weak interactions. 
This provides us with an insight into
the composition of dense matter which, of course, 
eventually needs to be benchmarked against observational constraints.

\begin{figure*}[t]
\begin{center}
  \includegraphics[width=1\linewidth]{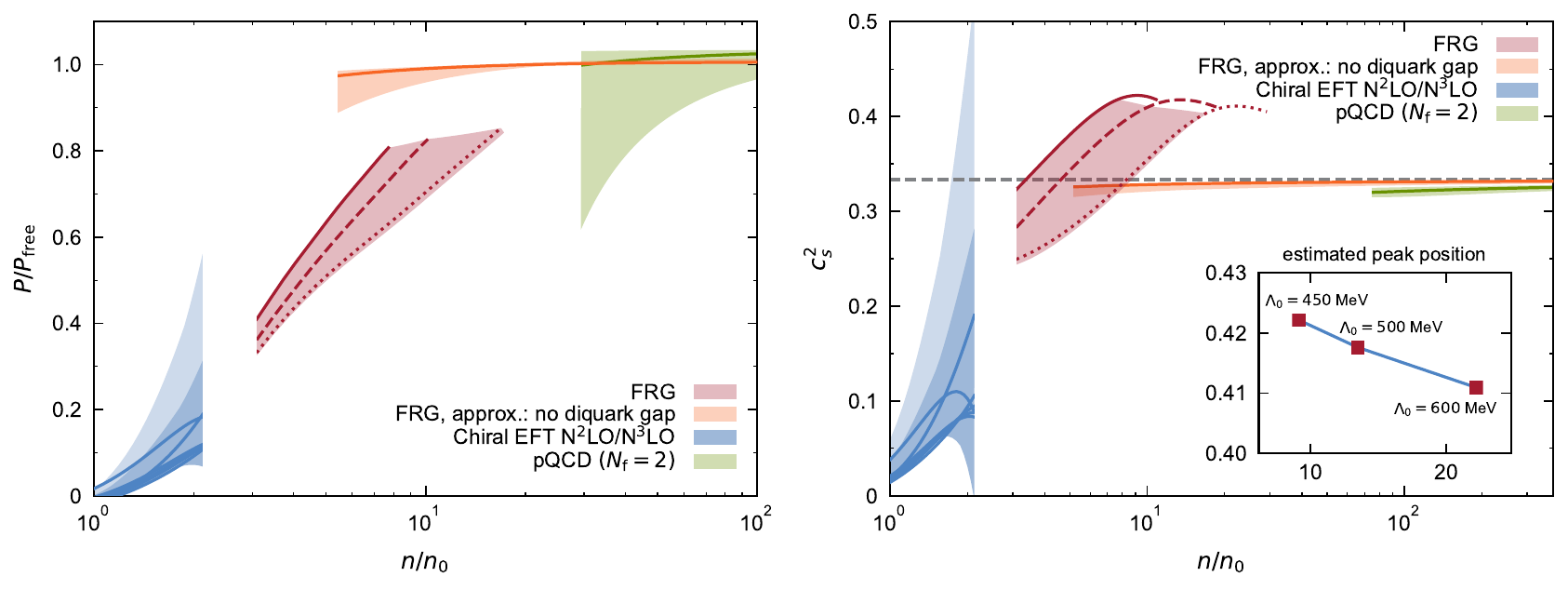}
\end{center}
\vspace*{-0.5cm}
\caption{Left panel: Pressure~$P$ of symmetric nuclear matter normalized by the pressure of the 
free quark gas~$P_\mathrm{SB}$ as a function of the baryon density~$n/n_0$ in units of the nuclear
saturation density as obtained from {chiral EFT, FRG, including results from an approximation without
taking into account a diquark gap (FRG, approx.: no diquark gap), and 
pQCD}, see main text for details. Right panel: Speed of sound squared as a
function of the baryon density in units of the nuclear saturation density as
derived from the pressure shown in the left panel. The inset shows the estimated peak position 
and height for different transition scales~$\Lambda_0$ as obtained by increasing 
the chemical potential~$\mu$ beyond~$\Lambda_0$.}
\label{fig:eos}
\end{figure*}

\begin{figure}[t]
\begin{center}
  \includegraphics[width=\linewidth]{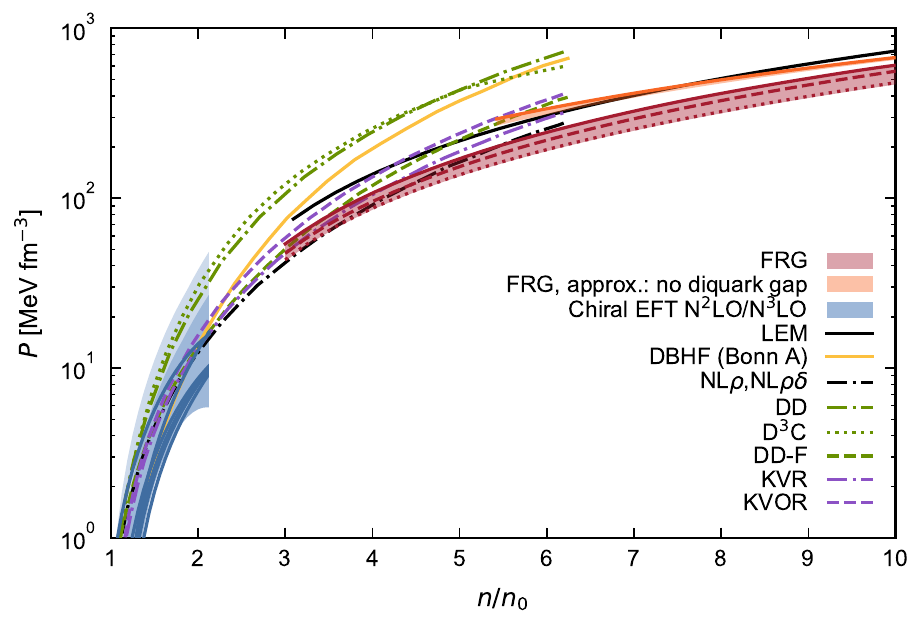}
\end{center}
\vspace*{-0.5cm}
\caption{Pressure of symmetric nuclear matter
as obtained from {chiral EFT,  FRG, and pQCD}, as in Fig.~\ref{fig:eos}, in comparison with different models
(see main text and also Ref.~\cite{Klahn:2006ir}).}
\label{fig:eos_comparison}
\end{figure}

{\it Low-density regime.--} 
At the fundamental level the
strong interaction is governed by the quark-gluon dynamics described by
quantum chromodynamics (QCD). At nuclear densities, the ground state is
dominated by chiral symmetry breaking and calculations directly based on QCD
become very challenging. For this regime chiral effective field theory (EFT)
represents a powerful framework to describe the nuclear dynamics and
interactions within a systematic expansion based on the low-energy
degrees of freedom, nucleons and
pions~\cite{Epelbaum:2008ga,Machleidt:2011zz}. Substantial progress has been
achieved in recent years in deriving new nuclear forces and computing the EOS
microscopically based on nucleon-nucleon (NN), three-nucleon (3N) and
four-nucleon (4N) interactions derived within chiral EFT
\cite{Hebeler:2009iv,Hebeler:2010xb,Tews:2012fj,Hagen:2013yba,Carbone:2013rca,Coraggio:2014nva,Wellenhofer:2015qba,Lynn:2015jua,Drischler:2015eba,Drischler:2017wtt}, 
see Ref.~\cite{Hebeler:2020ocj} for a recent review. 
In particular, in Ref.~\cite{Drischler:2017wtt} we presented an efficient
framework to compute the energy of nuclear matter at zero temperature within
many-body perturbation theory (MBPT) up to high orders in the many-body
expansion and for general proton fractions. It allows to include all
contributions from two- and many-body forces up to N$^3$LO and to explore
the{} connection of properties of matter and nuclei~\cite{Hoppe:2019uyw}. In
addition, in Ref.~\cite{Hebeler:2010xb} a set of NN and 3N interactions was
fitted to few-body observables, where all derived interactions led to good
saturation properties without adjustment of free parameters. In particular,
one interaction of this set was found to also correctly predict the ground
state energies of medium-mass nuclei up to $^{100}$Sn~\cite{Simonis:2017dny,Morris:2017vxi}. In
Figs.~\ref{fig:eos} and~\ref{fig:eos_comparison}, we show the results for the
pressure and the speed of sound of symmetric nuclear matter up to twice
nuclear density based on the set of interactions of Ref.~\cite{Hebeler:2010xb}
(individual blue lines) as well as the interactions up to N$^3$LO fitted to
the empirical saturation point of Ref.~\cite{Drischler:2017wtt} (blue bands).
The EFT uncertainty bands at N$^2$LO (light-blue band) and N$^3$LO (dark-blue
band) have been determined following the strategy of
Ref.~\cite{Epelbaum:2014efa} and represent the combined uncertainties based on
the results at the two cutoff scales $\Lambda = 450$ and $500$ MeV (see also
Ref.~\cite{Drischler:2017wtt}).

{\it Intermediate-density regime.--}
Compared to the nuclear density regime, less is known about the ground state at
intermediate densities, i.e., the regime above the region where calculations based on
chiral EFT are applicable and below the very high densities limit expected
to be governed, e.g., by the formation of a diquark
gap~\cite{Son:1998uk,Rapp:1997zu,Alford:1997zt,Berges:1998rc,Pisarski:1999tv,Pisarski:1999bf,Schafer:1999jg},
or {accessible by perturbative 
QCD (pQCD)} at asymptotic densities~\cite{Freedman:1976xs,Freedman:1976ub,Baluni:1977ms,Kurkela:2009gj,Gorda:2018gpy}.
To study the intermediate-density regime, we employ the {functional
renormalization group (FRG)} approach~\cite{Wetterich:1992yh} which allows us to
study this regime from the Euclidean QCD action (see
Refs.~\cite{Pawlowski:2005xe,Gies:2006wv,Braun:2011pp} for reviews):
\be
S = \int {\rm d}^4x\left\{ 
\frac{1}{4}F_{\mu\nu}^{a}F_{\mu\nu}^{a}
+ \bar{\psi}\left(
{\rm i}\slashed{\partial} + \bar{g}\slashed{A} +{\rm i}\gamma_0\mu \right)\psi
\right\}\,.
\label{eq:qcd}
\ee
Here,~$\bar{g}$ is the bare gauge coupling and~$\mu$ is the quark chemical
potential. The non-Abelian fields~$A_{\mu}^a$ enter the definition of the
field-strength tensor~$F_{\mu\nu}^a$ and are associated with the gluons. With
respect to the quarks, we restrict ourselves to two massless flavors in this
work.

In the RG flow, the quark-gluon vertex in Eq.~\eqref{eq:qcd} induces quark
self-interactions already at the one-loop level via two-gluon exchange. This
gives rise to terms, e.g., of the following form in the quantum effective
action:
\be
\delta\Gamma = \int {\rm d}^4x\,\bar{\lambda}_{i}(\bar{\psi}{\mathcal O}_i\psi)^2\,,
\label{eq:deltag}
\ee
where the operator~${\mathcal O}_i$ determines the color, flavor, and Dirac
structure of the vertex. We stress that the four-quark
couplings~$\bar{\lambda}_i$ are not free parameters but solely generated by
quark-gluon dynamics from first principles in our study. This is an important
difference to, e.g., Nambu-Jona-Lasinio-type model studies where the
four-quark couplings are input parameters, or to RG studies~\cite{Schafer:1998na,Son:1998uk} 
which expand the effective degrees of freedom around the Fermi surface in momentum space, 
thus being difficult to directly connect to the fundamental quark-gluon dynamics and the RG flow of the gauge coupling.  
In this work, we focus on the RG flow of pointlike projected four-quark
correlation functions~$\Gamma^{(4)}$. To be specific, we define the four-quark
couplings associated with the vertex of the form~\eqref{eq:deltag} as follows
\be
\label{eq:def4qc}
&&\bar{\lambda}_{i} (\bar{\psi}{\mathcal O}_i\psi)^2 \\
&&  =\! \lim_{p_i\to 0}
\bar\psi_{\alpha}(p_1) \bar\psi_{\beta}(p_2)
\Gamma^{(4), \alpha\beta\gamma\delta}_{\mathcal O}\! (p_1,p_2,p_3,p_4) \psi_{\gamma}(p_3) \psi_{\delta}(p_4).\nn
\ee
Here, $\alpha, \beta, \gamma, \delta$ denote collective indices for color,
flavor, and Dirac structures. Note that this zero-momentum projection does not
represent a Silver-Blaze-symmetric point~\cite{Khan:2015puu,Fu:2016tey,Braun:2017srn}, but
it matches the definition of four-quark couplings in conventional low-energy
models~\cite{Klevansky:1992qe,Buballa:2003qv,Fukushima:2011jc} and BCS-type
models~\cite{Bailin:1983bm,Altland:2006si,Alford:2007xm,Anglani:2013gfu}. The couplings resulting from our
definition~\eqref{eq:def4qc} depend on the chemical potential and the RG
scale. Although this scale dependence implies that part of the
momentum-dependent information is still taken into account in our RG analysis
in an effective manner~\cite{Braun:2014wja}, the pointlike limit ignores
relevant information of four-quark correlation functions. For example,
bound-state information is encoded in the momentum structure of these
correlation functions. Therefore, the pointlike approximation only allows us
to study the symmetric high-energy regime~\cite{Braun:2011pp}. The
symmetry-broken low-energy regime is not accessible in this way. For our
present purposes, however, this is still sufficient as it
enables us to study the approach towards the symmetry-broken low-energy regime, as
indicated by rapidly growing four-quark couplings.

In general, symmetry breaking is ultimately triggered by a specific four-quark
channel approaching criticality as indicated by a divergence of the
corresponding coupling. Such a seeming Landau-pole-type behavior of four-quark
couplings can be traced back to the formation of condensates as they can be
shown to be proportional to the inverse mass parameter of a Ginzburg-Landau
effective potential for the order parameters in a (partially) bosonized
formulation, $\lambda\sim 1/m^2$, see
Refs.~\cite{Ellwanger:1994wy,Gies:2001nw,Braun:2011pp}. On the one hand, this
implies that if the size of all
four-quark couplings is found to be bounded, the system stays in
the symmetric regime~\cite{Gies:2005as,Braun:2005uj,
Braun:2006jd,Braun:2011pp,Braun:2014wja}. On the other hand, the observation
of a rapidly growing four-quark coupling in a specific regime may be
considered as an indicator that the order-parameter potential develops a
non-zero ground-state expectation value in the direction associated with a specific
four-quark channel. The nontrivial assumption entering our analysis of the EOS
below is then that it is possible to relate the dominance pattern of the
four-quark couplings to the symmetry-breaking pattern in terms of condensates,
see Refs.~\cite{Braun:2014wja,Braun:2017srn,Braun:2018bik,2019arXiv190501060R} for a detailed
discussion. For example, in the zero-density limit, it has indeed been found
that the scalar-pseudoscalar channel is the most dominant
channel~\cite{Braun:2006zz,Mitter:2014wpa} and a corresponding condensate is
formed~\cite{Mitter:2014wpa,Cyrol:2017ewj} governing the low-energy dynamics.

When the baryon chemical potential is varied, it is reasonable to expect that
the symmetry-breaking patterns associated with the various four-quark channels
change. More specifically, channels other than the scalar-pseudoscalar
channel may become relevant. In general, the most dominant channel can be
identified by requiring that the modulus of the coupling of this channel is
greater than the ones of the other four-quark couplings. Such an analysis then
naturally requires to include all linearly-independent four-quark interactions
permitted by the $SU(\Nc)\otimes SU_\text{L}(2)\otimes SU_\text{R}(2)\otimes
U_\text{V}(1)$ symmetry. Taking into account the explicit breaking of
Poincar\'{e} and charge-conjugation invariance at finite density, we end up
with a Fierz-complete basis set composed of ten channels in the pointlike
limit~\cite{Braun:2018bik}. All other channels are related to this minimal
basis by means of Fierz transformations.

Introducing the dimensionless renormalized four-quark couplings $\lambda_i=k^2
\bar{\lambda}_i$ with~$k$ being the RG scale, the $\beta$ functions for these
couplings can be written in the following form:
\begin{equation}
k\partial_k\lambda_i =2\lambda_i - \lambda_j A^{(i)}_{jk} \lambda_k - B^{(i)}_j \lambda_j g^2 - C^{(i)} g^4, 
\label{eq:lambda}
\end{equation}
where~$i$ refers to the elements of our Fierz-complete basis of four-quark
couplings. The coefficients $A$ (purely fermionic loop), $B$ (triangle
diagram), and $C$ (two-gluon exchange) depend on the quark chemical potential.
Here, we have dropped an implicit dependence of these loop diagrams on the
wavefunction renormalization factors of the quarks and the gluons as they
have been found to be subleading in the symmetric
regime~\cite{Gies:2002hq,Gies:2003dp,Gies:2005as,Braun:2005uj, Braun:2006jd}.
For the computation of the flow equations~\eqref{eq:lambda}, we have then made
use of existing software packages~\cite{Huber:2011qr,Cyrol:2016zqb}. For
{details we refer the reader to Ref.~\cite{Braun:2019aow}. The} corresponding
flow equations for the purely fermionic part as parameterized by the
matrices~$A^{(i)}$ can be found in Ref.~\cite{Braun:2018bik}, including a
discussion of the regularization scheme also underlying this work.

In our present study, the RG flow of the gauge sector enters the flow
equations of the four-quark couplings only via the running of the strong
coupling. In line with our approximations in the computation of the four-quark
couplings, we only employ the one-loop running of the strong coupling for two
quark flavors. However, we have checked that our main results (e.g., the
existence of a maximum in the speed of sound) persist even if we employ
running couplings taking into account higher-order
effects~\cite{Braun:2005uj,Braun:2006jd,Braun:2014ata}. Note that from the
analysis of Ward-Takahashi identities, it follows that the back-reaction of
the four-quark couplings on the strong coupling is negligible, provided the
flow of the four-quark couplings is governed by the presence of fixed
points~\cite{Gies:2003dp}, as it is the case in the symmetric
regime~\cite{Gies:2005as,Braun:2005uj, Braun:2006jd}.

Using the set of flow equations defined by Eq.~\eqref{eq:lambda}, we can study
the RG flow of the four-quark couplings and analyze the emerging symmetry
breaking patterns. At high-energy scales, the RG flow generates quark
self-interactions~$\lambda_i\sim g^4$ via the last term in
Eq.~\eqref{eq:lambda}. Following the RG flow towards the low-energy regime, we
observe that the strength of the four-quark couplings relative to each other
depends on the dimensionless quark chemical potential~$\mu$. More
specifically, towards lower density, the most dominant channel in the low-energy
regime eventually turns out to be the scalar-pseudoscalar channel, in line
with phenomenological expectations. As also observed in
Ref.~\cite{Braun:2018bik}, the dominance pattern changes when the
dimensionless chemical potential~$\mu/k$ becomes sufficiently large. Then, the
diquark channel~$\sim\left( \I \bar \psi \gamma_5 \tau_2\, T^{A} \psi^C
\right) \left( \I \bar \psi^C \gamma_5 \tau_2\, T^{A} \psi \right)$ (where
$\tau_2$ is the second antisymmetric Pauli matrix and it is only summed over the
antisymmetric color generators~$T^{A}$) takes over the role of the most
dominant channel, suggesting the formation of a chirally symmetric diquark
condensate associated with pairing of the two-flavor color-superconductor (2SC) 
type~\cite{Rapp:1997zu,Alford:1997zt,Berges:1998rc,Pisarski:1999tv,Pisarski:1999bf,Schafer:1999jg}. 
Note that in case of electromagnetic neutrality and $\beta$ equilibrium the inclusion  
of strange quarks entails also different pairing patterns 
such as the color-flavor-locked pairing present at least at very high densities~\cite{Alford:1998mk}. 

For a computation of the EOS, it is required to solve the RG flow down to the
long-range limit~$k\to 0$. As discussed above, however, this requires to go
beyond the pointlike limit and resolve the momentum dependencies of the
corresponding vertices. For example, this can be conveniently done by
employing so-called dynamical hadronization
techniques~\cite{Gies:2001nw,Gies:2002hq,Pawlowski:2005xe,Floerchinger:2009uf},
see, e.g.,
Refs.~\cite{Braun:2008pi,Braun:2014ata,Mitter:2014wpa,Cyrol:2017ewj} for their
application to QCD. These techniques effectively implement continuous
Hubbard-Stratonovich transformations of four-quark interactions in the RG
flow. In the present work, we do not employ continuous transformations but
essentially perform them at a given scale~$\Lambda_0$~\cite{Springer:2016cji}.
To be specific, for any
given~$\mu$, we follow the RG flow of the four-quark couplings from the
perturbative high-energy regime down to the scale~$\Lambda_0$ at which we
extract the strength of the four-quark couplings and use them to fix the
couplings of an ansatz describing the dynamics at scales~$k<\Lambda_0$. Since
we find the scalar-pseudoscalar channel to be most dominant at low densities
and the diquark channel to be most dominant at intermediate and high
densities, in accordance with the findings in Ref.~\cite{Schafer:1998na},
we parametrize the low-energy regime associated with scales $k\leq
\Lambda_0$ by the Hubbard-Stratonovich transforms of these two couplings cast
into the form of a quark-meson-diquark-model truncation. From the latter, we
then compute the pressure via a minimization of the
corresponding Ginzburg-Landau-type effective potential~\cite{Fu:2018qsk} spanned by the
aforementioned two Hubbard-Stratonovich fields in an RG-consistent way, see
Ref.~\cite{Braun:2018svj} for details.

To set the scale, we fix the actual value of the scalar-pseudoscalar coupling
of the low-energy sector by the constituent quark mass in the vacuum limit.
The value of the diquark coupling relative to the scalar-pseudoscalar coupling
is then fixed by the corresponding ratio obtained from our RG flow study of
gluon-induced four-quark interactions evaluated at the scale~$\Lambda_0$.
Because the gluon-induced four-quark interactions depend on the quark chemical
potential, this renders the couplings of the low-energy regime
$\mu$-dependent. Finally, to estimate the uncertainties arising from the
presence of the scale~$\Lambda_0$ describing the ``transition" in the
effective degrees of freedom, we vary this scale from~$\Lambda_0=450\dots
600\,\text{MeV}$.

In the left panel of Fig.~\ref{fig:eos}, we show our results for the EOS
(light-red band) as a function of the baryon density in units of the nuclear
saturation density. The band has been obtained from a variation of the
scale~$\Lambda_0$ and a variation of the value of the gauge coupling within experimental 
errors at the initial RG scale~\cite{Bethke:2009jm}. The different line types within the light-red band
depict three representative EOSs associated with $\Lambda_0=450, 500,
600\,\text{MeV}$ (from left to right). At lower densities, we observe that our
results for the pressure as obtained from our many-body framework based on
chiral EFT interactions are remarkably consistent with those obtained from our
{FRG analysis} at intermediate densities. However, our present
approximation is not capable to resolve the exact position of any chiral
transition or crossover as we do not observe a clear dominance pattern in the spectrum
of the four-quark couplings in this regime. 
The extent of the light-red band
at high densities is set by the constraint~$\mu \leq
\Lambda_0$. 
{With respect to the high-density limit, we note that the results from our FRG studies are found to 
approach those from pQCD calculations (light-green band)~\cite{Kurkela:2009gj,Gorda:2018gpy}.}

In the right panel of Fig.~\ref{fig:eos}, we present the square of the speed
of sound as a function of the baryon density~$n$ as derived from the pressure
shown in the left panel. The light-red band is associated with the results
{from our FRG studies} taking diquark condensation into account. Its
extent to high densities is again constrained by the ``transition" scale~$\Lambda_0$. 
Irrespective of this limitation of our present study, 
a softening of the EOS at high densities may in general be expected from a 
perturbative standpoint as associated with an evaluation of, e.g., the four-quark 
couplings at the characteristic scale~$\mu\sim n^{\frac{1}{3}}$. In fact, at large chemical potential, 
the four-quark couplings~$\lambda_i$ then naturally become small owing to asymptotic freedom,~$\lambda_i\sim g^4$, see also Eq.~\eqref{eq:lambda}.
Even more, {according to pQCD} studies~\cite{Freedman:1976xs,Freedman:1976ub,Baluni:1977ms,Kurkela:2009gj,Gorda:2018gpy}, 
we expect the speed of sound to approach $c_{\rm s}^2=1/3$ (non-interacting limit) from below at asymptotically high densities.
From this and our observation
that the pressure exceeds its asymptotic value already for chemical potentials
well below the scale~$\Lambda_0$, we infer the existence of a maximum in the speed of sound. In order to also 
give an estimate for the position and height of the maximum, we may increase the
chemical potential~$\mu$ beyond~$\Lambda_0$ in our calculations. 
The inset of
this figure shows the resulting estimate, exhibiting a robust height of the
maximum but with a large uncertainty of its position. 
At high density, $n/n_0 > 75$, we show again results from
{pQCD calculations. Note} that the computation of the speed of sound
from the corresponding data for the pressure in this high-density branch
becomes numerically unstable for~$n/n_0\lesssim 70$.

{Besides pressure and speed of sound, the diquark gap~$\Delta$ is of great 
relevance for dense matter physics~\cite{Alford:2007xm}.
In our present study, we observe 
that~$\Delta$ increases within the considered density range~$n/n_0\approx 2\dots 15$, exhibiting 
a flattening towards higher densities. More specifically, we find that~$\Delta$ is greater than the values reported in 
Ref.~\cite{Alford:1997zt}. 
However, the values for~$\Delta$ from both studies show a remarkable agreement at lower densities.
For~$n/n_0 \approx 5$, for example, 
we extract $\Delta \approx 70\dots160\,\text{MeV}$ from Ref.~\cite{Alford:1997zt}  
and~$\Delta \approx 140\dots 230\,\text{MeV}$ (depending on~$\Lambda_0$) from our 
present study.}

In Fig.~\ref{fig:eos_comparison}, {we next compare our results for the pressure with different models.}
These include relativistic mean-field calculations, such as NL$\rho$ and
NL$\rho\delta$~\cite{Liu:2001iz}, DD, D${}^3$C and DD-F~\cite{Typel:2005b} as
well as KVR and KVOR~\cite{Kolomeitsev:2004ff} (see also
Ref.~\cite{Klahn:2006ir}). In addition, we show results of Dirac-Brueckner
Hartree-Fock calculations (DBHF)~\cite{vanDalen:2005sk} and from a
typical low-energy model (LEM)~\cite{Buballa:2003qv,Braun:2018svj}. At
densities up to around twice nuclear saturation density, the different models are
compatible with the chiral EFT uncertainty bands at N$^2$LO (but not all at N$^3$LO). At higher
densities, however, the pressure obtained from most models is
found to be significantly higher {than our results.}

{\it High-density regime.--} In the regime of very high densities the EOS can
be calculated using perturbative
methods~\cite{Freedman:1976xs,Freedman:1976ub,Baluni:1977ms,Kurkela:2009gj,Gorda:2018gpy}
owing to the fact that the dynamics is dominated by modes with
momenta~$|p|\sim\mu$ which effectively renders the QCD coupling~$g^2/4\pi$
small. Although the ground state is expected to be governed by diquark
condensation~\cite{Son:1998uk,Rapp:1997zu,Alford:1997zt,Pisarski:1999tv,Pisarski:1999bf,Schafer:1999jg},
calculations that do not include condensation effects are reliable,
provided that the chemical potential is much larger than the scale set by the
diquark~gap.

In our RG study, the gluon-induced four-quark interactions serve as proxies
for the various order parameters. The analysis of their RG flows indeed
indicate that the ground state is governed by spontaneous symmetry breaking,
even at high densities. This can be effectively described by a transition in
the relevant degrees of freedom at a finite scale. In order to make
contact with perturbative calculations, we drop the running of the four-quark
interactions and restrict ourselves to the running of the quark and gluon
wavefunction renormalization factors at leading order in the derivative
expansion. From the latter, we obtain dressed quark and gluon propagators
which are then used to compute the pressure. In this case, we find that the RG
flow of the pressure can be followed from high-energy scales down to the deep
infrared limit without encountering any pairing instabilities as associated
with spontaneous symmetry breaking. In Fig.~\ref{fig:eos}, we show our results
for the pressure and the speed of sound from this calculation labelled ``no diquark gap''.
We observe very
good agreement with recent perturbative calculations~\cite{Kurkela:2009gj,Gorda:2018gpy}.
The width of the orange FRG band illustrates the uncertainty arising from a variation of the
regularization scheme and a variation of the running gauge coupling within the
experimental error bars at the $\tau$-mass scale~\cite{Bethke:2009jm}.
Following the pressure towards smaller densities, we observe that our results
for the intermediate-density and high-density regime are consistent. For the
appearance of a maximum in the speed of sound, however, we find that the
inclusion of condensation effects in the regime~$n/n_0\lesssim 30$ is
crucial, which also provides the necessary decrease of the pressure in order 
to connect the low-density with the high-density regime.

{\it Conclusions and Outlook.--} In this Letter we have presented first
results for the EOS of symmetric nuclear matter at zero temperature over a
wide density range starting from QCD. At
low densities we performed calculations based on a set of recently developed
chiral NN and 3N interactions, while for densities beyond three times
saturation density we computed the EOS {within an FRG framework}
directly based on the fundamental quark-gluon dynamics. Even though the
present approximations underlying both studies break down at an
intermediate-density window, the results show a remarkable
consistency (in particular for the pressure) and indicate that they can be combined via simple extrapolations.
{At intermediate to high densities, our study suggests that the ground state is
governed by diquark dynamics. From a combined analysis of our results and those from
perturbative studies, we infer the existence of a maximum in the speed of sound.}
Although the exact position of this maximum in terms of the density 
cannot be determined conclusively in our present study, 
its height appears very robust. Note that 
the existence of a maximum for the speed of sound has also been 
demonstrated for neutron-rich matter based on constraints from 
neutron star masses~\cite{Bedaque:2014sqa,Tews:2018kmu,Greif:2018njt,Annala:2019puf}. 
Ignoring the diquark gap, {our FRG calculations} are then found to be
in good agreement with well-known results {from pQCD calculations}
at very high densities. A generalization of the presented framework to 
general proton fractions will give us access to the EOS in the neutron-rich regime,
which is relevant for astrophysical applications. Furthermore, {the
FRG approach} is already formulated for general temperatures. An
extension of our chiral EFT calculations at low densities to finite
temperatures will allow us to also study the temperature dependence of the EOS
over a wide density range based on strong interactions.
%

{\it Acknowledgments.--~} 
This work is supported in part by the Deutsche
Forschungsgemeinschaft (DFG, German Research Foundation) -- Projektnummer
279384907 -- SFB 1245,  the US Department of Energy, the Office of Science, the
Office of Nuclear Physics, and SciDAC under awards DE\_SC00046548 and
DE\_AC02\_05CH11231.
J.B. acknowledges support by the DFG under grant BR
4005/4-1 (Heisenberg program) and by HIC for FAIR within the LOEWE program of
the State of Hesse. 
C.D. acknowledges support by the Alexander von Humboldt Foundation through a Feodor-Lynen Fellowship.
Computational resources have been provided by the
Lichtenberg high performance computer of the TU Darmstadt. J.B., M.L., and
M.P. as members of the fQCD collaboration~\cite{fQCD} would like to thank the
other members of this collaboration for discussions and providing data for
cross-checks.


%
\bibliography{qcd}

\end{document}